\documentclass[aps,pra,preprint,superscriptaddress,amsmath]{revtex4}
\usepackage{longtable}
\usepackage{graphicx}
\usepackage{amssymb}
\usepackage{pstricks}

\newcommand{\e}{{\rm e}}
\newcommand{\drm}{{\rm d}}
\newcommand{\irm}{{\rm i}}
\newcommand{\beq}{\begin{equation}}
\newcommand{\eeq}{\end{equation}}
\newcommand{\bdm}{\begin{displaymath}}
\newcommand{\edm}{\end{displaymath}}
\newcommand{\bpsp}[2]{\begin{figure}[ht]\begin{center}\begin{pspicture}(#1,#2)}
\newcommand{\epsp}[2]{\end{pspicture}\caption{#1}\label{#2}\end{center}\end{figure}}

\begin{document}

\title{Quantum-Noise Power Spectrum of Fields with Discrete Classical Components}

\author{Jan Harms}
\author{Paul Cochrane}
\affiliation{Institut f\"ur Gravitationsphysik, Universit\"at Hannover and Max-Planck-Institut f\"ur Gravitationsphysik (Albert-Einstein-Institut), Callinstr.~38, 30167 Hannover, Germany}
\author{Andreas Freise}
\affiliation{School of Physics and Astronomy, University of Birmingham, Edgbaston, Birmingham B15 2TT, UK}


\date{\today}

\begin{abstract}
We present an algorithmic approach to calculate the quantum-noise spectral
density of photocurrents generated by optical fields with arbitrary discrete
classical spectrum in coherent or squeezed states. The measurement scheme
may include an arbitrary number of demodulations of the photocurrent.
Thereby, our method is applicable to the general heterodyne detection scheme
which is implemented in many experiments. For some of these experiments,
e.g.~in laser-interferometric gravitational-wave detectors, a reliable
prediction of the quantum noise of fields in coherent and squeezed states
plays a decisive role in the design phase and detector characterization.
Still, our investigation is limited in two ways. First, we only consider
coherent and squeezed states of the field and second, we demand that the
photocurrent depends linearly on the field's vacuum amplitudes which means
that at least one of the classical components is comparatively strong. 
\end{abstract}

\maketitle 

\section{Introduction}

Investigations of the quantum-noise spectrum usually elaborate on the
properties and correlations of the quantum vacuum while assuming simple
classical components of the field \cite{KLMTV01,BC01a,HSD04}. In another
publication the authors investigate the quantum-noise contribution to a
slightly more complex classical spectrum, i.e.~the outcome of a heterodyne
power measurement including one subsequent demodulation of the photocurrent
\cite{BCM02}. However, the classical spectrum of fields inside real
instruments is usually more complex, comprising many pairs of heterodyning
sidebands \cite{GEO06,SR07}. In addition, one may be interested in
measurement schemes which include more than one demodulation of the
photocurrent. In this paper, we extend previous analyses by allowing for an
arbitrary number of discrete classical components and an arbitrary number of
demodulations. Furthermore, our approach is algorithmic which means that the
results can be implemented straightforwardly in the code of simulators of
quantum-noise spectra.

In Sec.~\ref{secCoherent}, we introduce our notational conventions and
calculate spectral densities for power measurements of fields in coherent
states. The complexity is gradually increased, starting with a measurement
of discrete components without demodulation and ending with multi-component
fields including arbitrarily many demodulations. These results are further
generalized in Sec.~\ref{secSqueeze} to include fields in squeezed states.
For this purpose, the representation of the photocurrent is slightly but
essentially modified to take account of the intricate sideband correlations
occurring in squeezed fields. Finally in Sec.~\ref{secExample}, the
algorithm is applied to investigate the photocurrent noise spectrum for a
specific multi-component, squeezed field configuration which is meant to
clarify the abstract approach outlined in previous sections.

\section{Quantum-Noise Spectrum of Coherent Fields}
\label{secCoherent}

Let us first calculate the quantum-noise spectral density of a coherent
field which is determined by a single classical amplitude $c_0$ at frequency
$f_0$. Denoting the quantum vacuum noise amplitudes by $\hat{q}(f)$, the
electric field can be written
\beq
\begin{split}
\hat{E}(t)&=c_0\e^{-2\pi\irm f_0t}+c_0^*\e^{2\pi\irm f_0t}\\
&+\int\limits_0^\infty\drm f\left(\hat{q}(f)\e^{-2\pi\irm ft}+\hat{q}^\dagger(f)\e^{2\pi\irm ft}\right).
\end{split}
\eeq
Factoring out the oscillating phase, $\exp(-2\pi\irm f_0t)$, of the classical
field, the quantum vacuum integration is carried out over sideband
frequencies $F=f-f_0$. The range of sideband frequencies is restricted since
we consider a limited measurement bandwidth: $F\in[-B,B]$ ($B\ll f_0$). We
also split the field into two Hermitian conjugate parts $\hat{E}^{(+)}(t)$,
$\hat{E}^{(-)}(t)$ and so the positive-frequency field is
\beq
\hat{E}^{(+)}(t)=\e^{-2\pi\irm f_0t}\left(c_0+\int\limits_{-B}^B\drm
F\hat{q}(f_0+F)\e^{-2\pi\irm Ft}\right).
\label{eqSingle}
\eeq
When talking of spectral densities, we mean the power spectral density
of the photocurrent after demodulations. Ideally, the photocurrent
$\hat{I}(t)$ is proportional to the power of the field. For real
photodiodes, this does not have to be true for arbitrarily large detection
bandwidths. In the simplest case the current spectral density has to be multiplied with a factor which accounts for the frequency dependent response of the photodiode. Let us assume that $B$ is small enough so that $\hat{I}\propto\hat{P}$ for all measured frequencies. The power of the field Eq.~(\ref{eqSingle}) averaged over a period $\tau\ll c/B$ is given by
\beq
\begin{split}
\hat{P}(t)&=\hat{E}^{(-)}(t)\hat{E}^{(+)}(t)\\
&=\frac{1}{2}|c_0|^2+\int\limits_{-B}^B\drm Fc_0^*\hat{q}(f_0+F)\e^{-2\pi\irm Ft}+{\rm h.c.}
\end{split}
\eeq
Contributions quadratic in the vacuum amplitudes are neglected which is a
valid approximation whenever $P_0\equiv|c_0|^2\gg hf_0/\tau$ , where $\tau$
denotes the measurement time. It follows that the optical quantum noise of
the photocurrent $\hat{I}^{\rm QM}(t)$ is proportional to
\beq
\hat{I}^{\rm QM}(t)\propto \int\limits_{-B}^B\drm Fc_0^*\hat{q}(f_0+F)\e^{-2\pi\irm Ft}+{\rm h.c.}
\label{eqCurrQM}
\eeq
The proportionality factor $\chi$ between current and light power---the
photodiode responsivity---typically assumes values $\chi=0.2-0.7\, \rm A/W$
at $f\sim 3\cdot 10^{14}\,\rm Hz$ \cite{Rak01}. Since subsequent
calculations are based on a frequency independent responsivity, there is no
real distinction between light-power and photocurrent power spectral
densities except for a factor of $\chi^2$. Throughout this paper the explicit
dependence of the photocurrent on $\chi$ is omitted.

Vacuum amplitudes like any classical amplitudes of stationary noise are
numerically ill-defined (the Fourier transform of stationary noise being
infinite) and serve exclusively as algebraically meaningful quantities.
Nevertheless, the power spectral density of stationary noise is well-defined
and there exists a simple relationship between stationary vacuum noise
amplitudes $\hat q$ and its (single-sided) noise spectral densities $S_{\rm
q}$
\beq
\begin{split}
\frac{1}{2}\delta(f-f^\prime)\,S_{\rm q}(f)&=\frac{1}{2}\langle\hat{q}(f)\hat{q}^\dagger(f^\prime)+\hat{q}^\dagger(f^\prime)\hat{q}(f)\rangle\\
&=\frac{1}{2}\delta(f-f^\prime)\,hf.
\end{split}
\eeq
The corresponding quantum noise spectral density of the photocurrent
Eq.~(\ref{eqCurrQM}) which results from the power measurement of a coherent
field reads \cite{CS85}
\beq
S^{\rm QM}_{\rm I}(F)=|c_0|^2\cdot S_{\rm q}(f_0)=P_0\cdot h f_0.
\label{eqSpecSimple}
\eeq
The invalidity of the latter equation for squeezed states of the field and
the fact that the spectral density does not depend on the sideband frequency
$F$ is clarified in Sec.~\ref{secSqueeze} (see Eq.~(\ref{eqSpecTwo})). At
this point, we only wish to draw attention to the apparently simple
algorithm for coherent fields which leads to Eq.~(\ref{eqSpecSimple})
starting from Eq.~(\ref{eqSingle}): square the classical amplitude and
multiply the result by $hf_0$. 

In fact, the algorithm stays valid for a wider class of power measurements.
Consider a set of classical components with amplitudes $c_{\rm i}\equiv
c(f_{\rm i})$, $i\in\{0,1,\ldots,N-1\}$. Like before, we factor out the
phase of the component $c_0$ which formally converts all optical
oscillations into sideband oscillations with respect to the reference
frequency $f_0$. Defining $F_{\rm i}\equiv f_{\rm i}-f_0$, the
positive-frequency field is cast into the form
\beq
\hat{E}^{(+)}(t)=\e^{-2\pi\irm f_0t}\sum\limits_{i=0}^{N-1}\bigg(c_i\e^{-2\pi\irm F_it}+\int\limits_{F_i-B}^{F_i+B}\drm F\hat{q}(f_0+F)\e^{-2\pi\irm Ft}\bigg).
\eeq
We demand that $P_{\rm i}\equiv|c_{\rm i}|^2\gg hf_0/\tau$ is valid for all
classical components. If that condition did not hold for some classical
components (typically for signal sidebands), then the respective amplitudes
would not influence the quantum-noise power spectrum as long as at least one
high power component exists and so for our purpose we may safely neglect
weak components. A power measurement yields a
photocurrent whose quantum noise is determined by
\beq
\begin{split}
\hat{I}^{\rm QM}(t)&=\sum\limits_{i=0}^{N-1}\int\limits_{F_i-B}^{F_i+B}\drm Fc_i^*\hat{q}(f_0+F)\e^{-2\pi\irm (F-F_i)t}+{\rm h.c.}\\
&=\sum\limits_{i=0}^{N-1}\int\limits_{-B}^B\drm Fc_i^*\hat{q}(f_0+F+F_i)\e^{-2\pi\irm Ft}+{\rm h.c.}
\label{eqCurrCohMult}
\end{split}
\eeq
Provided that all intervals $[F_i-B,F_i+B]$ are mutually disjoint, the
quantum-noise power spectral density of the photocurrent is given by
\beq
S^{\rm QM}_{\rm I}(F)=\sum\limits_{i=0}^{N-1}|c_i|^2\cdot S_{\rm q}(f_0+F_i)=\sum\limits_{i=0}^{N-1} h\cdot(f_0+F_i)\cdot P_i
\label{eqSpecCohMult}
\eeq
Unfortunately, there are no more simple example cases which could be
presented. Let us now skip to the next section and introduce demodulations of
photocurrents.

\subsection{A Single Demodulation}

There are two well-known demodulation techniques. Either the photocurrent is
multiplied by a harmonic function $\hat{I}(t)\cdot\cos(2\pi D\cdot
t+\phi)$ or one rectifies the photocurrent which means the final output
is $|\hat{I}(t)|$. In practice, the low-frequency spectra ($B\ll
F_{\rm i}$) drawn from these two outputs differ by a constant factor, the
spectrum of the harmonic demodulation being smaller by a factor of $(\pi/4)^2$.
Anyway, the spectrum of the harmonically demodulated current is calculable by
much simpler algebra. In addition, we want to study measurement schemes
which implement multiple demodulations characterized by a set of
demodulation frequencies $ D_{\rm i}$ and demodulation phases $\phi_{\rm
i}$. For these two reasons, we consider harmonic demodulations throughout
this paper. Let us perform a single demodulation of the current
Eq.~(\ref{eqCurrCohMult})
\beq
\begin{split}
\hat{I}^{\rm QM}(t)&=\frac{1}{2}\sum\limits_{i=0}^{N-1}\int\limits_{-D-B}^{- D+B}\drm Fc_i^*\hat{q}(f_0+F+F_i)\e^{-\irm(2\pi(F+D)t+\phi)}\\
&\quad+\frac{1}{2}\sum\limits_{i=0}^{N-1}\int\limits_{D-B}^{D+B}\drm Fc_i^*\hat{q}(f_0+F+F_i)\e^{-\irm(2\pi(F-D)t-\phi)}\\
&\quad+\rm h.c.
\end{split}
\eeq
or shifting the frequency range, the current noise finally assumes the form
\beq
\hat{I}^{\rm QM}(t)=\dfrac{1}{2}\sum\limits_{i=0}^{N-1}\int\limits_{-B}^B\drm F\e^{-2\pi\irm Ft}\Big(c_i^*\e^{-\irm\phi}\hat{q}(f_0+F+F_i- D)+c_i^*\e^{\irm\phi}\hat{q}(f_0+F+F_i+ D)\Big)+\rm h.c.
\label{eqCurrCoh2c1d}
\eeq
Now, if all frequency intervals $[F_i- D-B,F_i- D+B],\,[F_i+ D-B,F_i+ D+B]$
are mutually disjoint, then one obtains exactly the same spectral density as
in Eq.~(\ref{eqSpecCohMult}). However, this situation is uncommon in real
experiments. 

It is time to introduce a graphical auxiliary in order to understand what is
happening. For simplicity we start with two classical components----a carrier 
and a subcarrier---at
frequencies $F_0=0\,\rm Hz$ and $F_1=30\,\rm MHz$. The photocurrent is
demodulated with $ D=15\,\rm MHz$ and the detection bandwidth is limited to
$B=1000\,\rm Hz$. This choice of frequencies also corresponds to a very common
situation in optical experiments, namely the heterodyne measurement where
the two classical components are generated by a $15\,\rm MHz$ modulation of
a carrier field which oscillates at high optical frequencies ($f_0\sim
10^{15}\,\rm Hz$). We point out that $B\ll
|F_0\pm D|,|F_1\pm D|,|F_0\pm F_1|$ which significantly simplifies the
problem. The four frequency values $F_{00}\equiv F_0- D,\,F_{01}\equiv F_0+
D,\,F_{10}\equiv F_1- D$ and $F_{11}\equiv F_1+ D$ are marked on a frequency
axis (see Fig.~\ref{figFA2c1d}) and collected inside a matrix $\mathcal{F}(N=2,2)=\{F_{00},\,F_{01};\; F_{10},\,F_{11}\}$.
\bpsp{7.5}{2}
\psline{->}(0,1)(7,1)
\psline(2,0.95)(2,1.05)
\psline(5,0.95)(5,1.05)
\psframe[fillstyle=solid,fillcolor=black](0.3,0.9)(0.7,1.1)
\psframe[fillstyle=solid,fillcolor=black](3.3,0.9)(3.7,1.1)
\psframe[fillstyle=solid,fillcolor=black](6.3,0.9)(6.7,1.1)
\put(0.3,0.6){$F_{00}$}
\put(1.9,1.2){$F_0$}
\put(2.8,0.6){$F_{01}=F_{10}$}
\put(4.9,1.2){$F_1$}
\put(6.3,0.6){$F_{11}$}
\put(7.1,0.9){$F$}
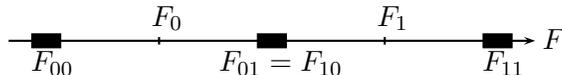
\epsp{Representation of the relevant frequency ranges for a classical heterodyne power measurement. The black boxes indicate the detection bandwidth.}{figFA2c1d}
Since two intervals identically overlap, each vacuum amplitude at
frequencies inside that interval enters twice into the spectral density
calculation. These two contributions have to be added coherently before
taking the absolute square. The expression Eq.~(\ref{eqCurrCoh2c1d}) for the
current tells us that all vacuum amplitudes $\hat{q}(f_0+F+15\,\rm MHz)$
have to be added coherently. In conclusion, the spectral density in this
particular example turns out to be
\beq
\begin{split}
S^{\rm QM}_{\rm I}(F)&=\frac{1}{4}\big(h\cdot(f_0-15\,{\rm MHz})\cdot P_0+h\cdot(f_0+15\,{\rm MHz})\cdot|c_0^*\e^{-\irm\phi}+c_1^*\e^{\irm\phi}|^2\\
&\qquad+h\cdot(f_0+45\,{\rm MHz})\cdot P_1\big)\\
&\approx\frac{hf_0}{4}\left(P_0+P_1+|c_0^*\e^{-\irm\phi}+c_1^*\e^{\irm\phi}|^2\right).
\end{split}
\label{eqSpecCoh2c1d}
\eeq
Setting $c_1=0$, one may wonder where half the noise power has gone when
comparing the result with Eq.~(\ref{eqSpecSimple}). Basically, the answer is
that by demodulating we multiply the photocurrent with a function which, in terms of power, has
a gain of 1/2. 

Henceforth, we will always assume that overlapping frequency ranges represented by black boxes in Fig.~\ref{figFA2c1d} overlap
completely, but never partially. The overlap condition guarantees
that whenever two frequencies inside the frequency matrix $\mathcal{F}$ do
not coincide, then the two respective detection ranges do not overlap and
one does not have to worry about coherent summation of the respective
amplitudes. This is certainly a reasonable demand for a first approach to an algorithmic
realization of the calculation. 

A discussion of the result Eq.~(\ref{eqSpecCoh2c1d}) focussing on
signal-to-noise ratios can be found in \cite{BCM02}. We are now prepared to
calculate the spectral density for measurements with a single demodulation
and an arbitrary number of classical components. It is not possible to give
explicit results, because as we have seen these depend on the chosen set of
field and demodulation frequencies. Our focus lies on extending the
algorithm which leads to the spectral density. Consider $N$ classical
components $c_{\rm i}$ at frequencies $F_{\rm i}$ and a single demodulation
of the photocurrent. The first step is to calculate the frequency matrix
\beq
\mathcal{F}(N,2)=
\begin{pmatrix}
F_{00} & F_{01}\\
F_{10} & F_{11}\\
\vdots & \vdots\\
F_{N-1,0} & F_{N-1,1}
\end{pmatrix}
\label{eqMatrix}
\eeq
where $F_{\rm i0}\equiv F_{\rm i}- D$ and $F_{\rm i1}\equiv F_{\rm i}+ D$.
The second step is to find coinciding frequencies of the matrix. Use the
matrix indices of these pairs, e.g. $(n_1,d_1)$ and $(n_2,d_2)$, to
calculate the contribution to the spectral density as follows
\beq
S^{\rm QM}_{\rm I}(F,(n_1,d_1),(n_2,d_2))=\frac{h\cdot(f_0+F_{n_1d_1})}{4}\cdot|c_{n_1}^*\e^{(-1)^{d_1+1}\irm\phi}+c_{n_2}^*\e^{(-1)^{d_2+1}\irm\phi}|^2
\label{eqSpecNc1d}
\eeq
All remaining unique frequencies $F_{\rm nd}$ contribute according to
\beq
S^{\rm QM}_{\rm I}(F,(n,d))=\frac{h\cdot(f_0+F_{nd})}{4}|c_n|^2.
\label{eqSpecUni}
\eeq
In most experiments $f_0\gg F_{\rm nd}$ and the quantum vacuum energies can
be approximated by $hf_0$. Finally, one has to sum up all these
contributions. 

Let us summarize our preliminary results as a list to be processed when 
calculating the current noise spectral density:
\begin{enumerate}
\item Calculate the matrix $\mathcal{F}$ according to Eq.~(\ref{eqMatrix}).
\item Collect index pairs ($n,d$) of coinciding frequencies $F_{\rm nd}$
inside $\mathcal F$.
\item Collect contributions to the current noise spectral density from
unique frequencies which are determined by Eq.~(\ref{eqSpecUni}).
\item Collect contributions to the current noise spectral density from
coinciding frequencies which are determined by Eq.~(\ref{eqSpecNc1d}).
\item Sum up all contributions.
\end{enumerate}
This algorithm needs an extension in order to account for
multiple demodulations. This task is accomplished in the next section.

\subsection{$M$ Demodulations}

Methodically, increasing the number of demodulations means to increase the
number or range of indices in the previous calculations. The first step of
the algorithm is always to collect all relevant frequencies inside a matrix.
Again, we consider a set of $N$ classical components $c_{\rm i}$ at
frequencies $F_{\rm i}$. The demodulations are determined by $M$
demodulation frequencies $ D_{\rm i}$ and phases $\phi_{\rm i}$. The final
output of the measurement is given by
\beq
\hat{I}(t)\cdot\prod\limits_{i=0}^{M-1}\cos(2\pi D_i\cdot t+\phi_i)=\hat{I}(t)\cdot\frac{1}{2^M}\sum\limits_{i=1}^{2^M}\cos\left(\sum\limits_{j=0}^{M-1}(-1)^{\lfloor i/2^{M-j-1}\rfloor}\left(2\pi D_j\cdot t+\phi_j\right)\right)
\label{eqMultDemod}
\eeq
with $\lfloor x\rfloor$ denoting the decimal truncating floor function.
Evaluating Eq.~(\ref{eqMultDemod}) for a few small number of demodulations,
one recognizes that the arguments of the harmonic functions consist of all
possible $\pm$-combinations of modulation frequencies and phases. Since the
complete sum of these harmonic functions is multiplied with the
photocurrent, it should be obvious that each row $i$ of the frequency matrix
$\mathcal F$ contains all frequencies of the form $F_{\rm i}\pm
D_0\pm\ldots\pm D_{M-1}$. In which order should these $2^M$ frequencies
appear? One may argue that the order does not matter in principle. However,
a specific combination of frequencies is associated with an analogous
combination of demodulation phases which then appears in formulas like
Eq.~(\ref{eqSpecNc1d}). If two frequencies of $\mathcal{F}$ turn out to be
equal, then the two pairs of matrix indices have to encode the
$\pm$-combination which determines the respective combination of phases.
Algorithmic tractability of the problem requires a good sorting scheme of
these frequencies. We propose a sorting scheme which is derived from a tree
structure like that in Fig.~\ref{figTree}.
\bpsp{8}{4}
\psline(4,3.5)(4,3)
\psline(2,2.2)(2,3)(6,3)(6,2.2)
\psline(1,1.4)(1,2.2)(3,2.2)(3,1.4)
\psline(5,1.4)(5,2.2)(7,2.2)(7,1.4)
\psline(0.4,0.6)(0.4,1.4)(1.6,1.4)(1.6,0.6)
\psline(2.4,0.6)(2.4,1.4)(3.6,1.4)(3.6,0.6)
\psline(4.4,0.6)(4.4,1.4)(5.6,1.4)(5.6,0.6)
\psline(6.4,0.6)(6.4,1.4)(7.6,1.4)(7.6,0.6)
\put(3.9,3.6){$F_{\rm i}$}
\put(2.4,3.1){$-D_0$}\put(5,3.1){$+D_0$}
\put(0.9,2.3){$-D_1$}\put(2.4,2.3){$+D_1$}\put(4.9,2.3){$-D_1$}\put(6.4,2.3){$+D_1$}
\put(0.2,1.5){$-D_2$}\put(1.1,1.5){$+D_2$}\put(2.2,1.5){$-D_2$}\put(3.1,1.5){$+D_2$}
\put(4.2,1.5){$-D_2$}\put(5.1,1.5){$+D_2$}\put(6.2,1.5){$-D_2$}\put(7.1,1.5){$+D_2$}
\put(0.2,0.3){$F_{\rm i0}$}\put(1.4,0.3){$F_{\rm i1}$}\put(2.2,0.3){$F_{\rm i2}$}\put(3.4,0.3){$F_{\rm i3}$}
\put(4.2,0.3){$F_{\rm i4}$}\put(5.4,0.3){$F_{\rm i5}$}\put(6.2,0.3){$F_{\rm i6}$}\put(7.4,0.3){$F_{\rm i7}$}
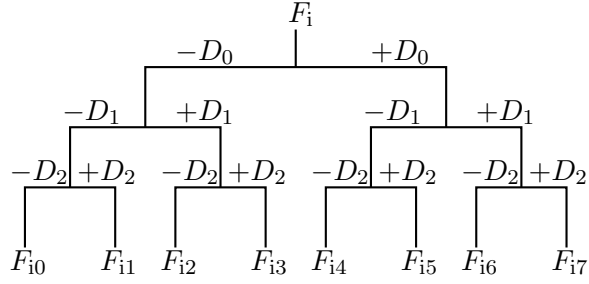
\epsp{Sorting scheme for three demodulation frequencies.}{figTree}
The frequencies themselves do not obey $F_{\rm i0}<F_{\rm i1}<\ldots$ since we
do not assume a magnitude-sorted vector of demodulation frequencies. Now,
the remaining problem is to derive from each column index the respective
combination of demodulation phases. If each row vector of $\mathcal F$ is
tree sorted then one first converts the decimal column index $j$ into a
binary number $(j)_{\rm bin}=[b_0b_1b_2b_3\ldots b_{M-1}]$ and then calculates
a total phase according to
\beq
\phi(j)=-\sum\limits_{i=0}^{M-1}(-1)^{b_i}\phi_i.
\label{eqPhase}
\eeq
This phase is to be used in order to calculate contributions to the quantum
noise spectral density when $C$ frequencies of $\mathcal F$ coincide. If the
respective pairs of indices are $(n_{\rm j},d_{\rm j})$, $j\in\{0,..,C-1\}$,
then the spectral density reads
\beq
S^{\rm QM}_{\rm I}(F,\{(n_{\rm j},d_{\rm j})\})=\frac{h\cdot(f_0+F_{n_0d_0})}{4^M}\left|\sum\limits_{j=0}^{C-1}c_{n_j}^*\e^{\irm\phi(d_j)}\right|^2.
\label{eqSpecNcMd1}
\eeq
The contribution from unique frequencies $F_{\rm nd}$ is given by
\beq
S^{\rm QM}_{\rm I}(F,(n,d))=\frac{h\cdot(f_0+F_{nd})}{4^M}|c_n|^2.
\label{eqSpecNcMd2}
\eeq
Finally, let us apply the algorithm to the case with $N=2$ classical
components and $M=3$ demodulations. The respective frequency matrix is
written
\beq
\mathcal{F}\left(2,8\right)=
\begin{pmatrix}
F_{00} & F_{01} & F_{02} & F_{03} & F_{04} & F_{05} & F_{06} & F_{07}\\
F_{10} & F_{11} & F_{12} & F_{13} & F_{14} & F_{15} & F_{16} & F_{17}
\end{pmatrix}.
\eeq
Coincidences are found say between frequencies $F_{05}$ and $F_{11}$. The
binary representation of the first column index is $(5)_{\rm bin}=101$ and
$(1)_{\rm bin}=001$ for the second. According to Eq.~(\ref{eqPhase}),
one obtains $\phi(5)=\phi_0-\phi_1+\phi_2$ and
$\phi(1)=-\phi_0-\phi_1+\phi_2$. The respective spectral density is given by
\beq
S^{\rm QM}_{\rm I}(F,(0,5),(1,1))=\frac{h\cdot(f_0+F_{11})}{64}\left|c_0^*\e^{\irm\phi(5)}+c_1^*\e^{\irm\phi(1)}\right|^2.
\eeq
Contributions coming from unique frequencies are calculated as usual. The
first part of this paper is finished. We have treated the calculation of
quantum noise when the field is coherent which entails that the noise is time
stationary. In fact, the algorithm is valid for any type of time-stationary
noise including technical laser noise when phase and amplitude noise are not
correlated. In that case, one has to substitute the quantum vacuum energies
in Eqs.~(\ref{eqSpecNcMd1})\&(\ref{eqSpecNcMd2}) by another noise spectral density which
characterizes the power of the technical noise at the photodiode. The next
step is to take into account the intricate correlation between quantum noise
amplitudes at different frequencies due to squeezing or in other words, due
to amplitude-phase correlations.

\section{Quantum-Noise Spectrum of Squeezed Fields}
\label{secSqueeze}

A widely applied mechanism which correlates vacuum noise amplitudes at
different frequencies is the generation of \emph{squeezed fields}.
Squeezed fields are formed in nonlinear crystals \cite{Va06} and theory
predicts that ponderomotive squeezing occurs when light is reflected from
suspended mirrors \cite{BC01b}. In order
to understand the meaning of squeezing, one has to know that
correlations are built up between sideband frequencies $F$ with respect to a
reference frequency $f_{\rm i}$. The squeezing transformation of fields is
characterized by a squeezing factor $r$ which quantifies the strength of
correlations between different frequencies and a squeezing phase $\phi$.
Perfect correlation corresponds to a squeezing factor $r=\infty$.  In
practice, the reference frequency is realized by
means of a single classical component which is named seed or carrier field
depending on whether the squeezing is generated by crystals or
ponderomotively. Now, if $\phi=0$ one can show that the correlation
between sidebands diminishes the quantum amplitude noise of the classical
field and for $\phi=\pi/2$, the quantum phase noise is
decreased. Interpretation of the squeezing phase for intermediate values
requires a more sophisticated representation of fields \cite{CS85}. In
principle, the reference frequency does not have to be related to a
classical field. In that case, the squeezing phase has no {\it ad hoc}
interpretation. Denoting amplitudes of squeezed fields by $\hat{s}$ and
amplitudes of coherent fields by $\hat q$, the squeezing transformation is
governed by \cite{CS85}
\beq
\hat{s}(f_{\rm i}\pm F)=\cosh(r)\cdot \hat{q}(f_\irm\pm F)+\sinh(r)\e^{2\irm\phi}\cdot \hat{q}^\dagger(f_\irm\mp F).
\label{eqSqueeze}
\eeq
How does this squeezing transformation relate to our previous investigation
of quantum noise spectral densities? To find an answer, we have to return
to Eq.~(\ref{eqCurrCohMult}). That equation determines the time-dependent
photocurrent where $F$, the former optical sideband frequencies, now become
true frequencies of the current spectrum. Similarly to expansions of
electric fields, we had better conform to a strict decomposition into
positive and negative frequencies
\beq
\hat{I}^{\rm QM}(t)=\sum\limits_{i=0}^{N-1}\int\limits_0^B\drm F\e^{-2\pi\irm F\cdot t}\Big(c_i^*\hat{s}(f_0+F+F_i)+c_i\hat{s}^\dagger(f_0-F+F_i)\Big)+{\rm h.c.}
\label{eqCurrPN}
\eeq
where the vacuum amplitudes have been renamed to indicate the possibility of squeezing. Thereby, the impact of
Eq.~(\ref{eqSqueeze}) on the power spectral density of the photocurrent
becomes clear. Two amplitudes at different frequencies of the optical vacuum
are added to form a single amplitude of the photocurrent. If these two
amplitudes are uncorrelated then the spectral density is calculated as
before making implicit use of the identity 
\beq
S(\hat q_1+\hat q_2)=S(\hat q_1)+S(\hat q_2), \qquad \mbox{$\hat q_1,\,\hat q_2$ uncorrelated.}
\eeq 
The optical field may exhibit correlations due to squeezing and the latter
equation can no longer be applied. So the remaining problem is the
calculation of the modified spectral density depending on the squeezing
factor and phase. There exist two possibilities. Either squeezing is
generated whose reference frequency coincides with one of the frequencies
$f_0+F_\irm$ which appear in the current expansion Eq.~(\ref{eqCurrPN}) and
the squeezing factor is significant only within a frequency range comparable
to the detection bandwidth $B$, or the squeezing violates either of these two
conditions. We start with an investigation of the case when both conditions
are fulfilled which is in some sense the only expedient one. It is then also
straightforward to treat fields which are squeezed at some or all of the
frequencies $f_0+F_\irm$ with potentially different factors and phases. The
correction is derived from the quantum-noise amplitude of the photocurrent;
\beq
\hat p(F,c_\irm^*,F_\irm)\equiv c_\irm^*\hat{s}(f_0+F+F_\irm)+c_\irm\hat{s}^\dagger(f_0-F+F_\irm).
\eeq
Before or after squeezing, the field may be subject to linear (frequency
preserving) transformations like propagations or reflections from fixed
mirrors. If some mirrors are suspended then already squeezed fields
experience further squeezing. Consequently, a more generic case is
considered here described by the following pair of transformations:
\beq
\begin{split}
\hat s(f_\irm+F)&=t_{00}(f_\irm+F)\hat q(f_\irm+F)+t_{01}(f_\irm-F)\hat q^\dagger(f_\irm-F)\\
\hat s^\dagger(f_\irm-F)&=t_{10}(f_\irm+F)\hat q(f_\irm+F)+t_{11}(f_\irm-F)\hat q^\dagger(f_\irm-F).
\end{split}
\label{eqTransform}
\eeq
The two transfer functions $t_{10},\,t_{01}$ map input amplitudes to output
amplitudes at the photodiode with mutually different frequencies, whereas 
$t_{00},\,t_{11}$ comprise nonlinear and linear transfers as mentioned above. 
The input vacuum is coherent and therefore all amplitudes $\hat q$ of the
input vacuum are uncorrelated. Inserting the last equation into the
amplitude of the photocurrent and defining $t_{\rm mn}^\irm(\pm)\equiv
t_{\rm mn}(f_0\pm F+F_\irm)$, one obtains
\beq
\begin{split}
\hat p(F,c_\irm^*,F_\irm)&= [c_i^*t_{00}^\irm(+)+c_it_{10}^\irm(+)]\cdot\hat{q}(f_0+F+F_i)\\
&\quad+[c_i^*t_{01}^\irm(-)+c_it_{11}^\irm(-)]\cdot\hat{q}^\dagger(f_0-F+F_i).
\end{split}
\label{eqCurrAmp}
\eeq
The power spectral density associated with the noise amplitude $\hat
p(F,c_\irm^*,F_\irm)$ is given by
\beq
\begin{split}
S(\hat p(F,c_\irm^*,F_i))&=\frac{h\cdot(f_0+F_\irm)}{2}\Bigg[|c_i^*t_{00}^\irm(+)+c_it_{10}^\irm(+)|^2\cdot \left(1+\frac{F}{f_0+F_i}\right) \\
&\qquad\qquad\qquad\quad+|c_i^*t_{01}^\irm(-)+c_it_{11}^\irm(-)|^2\cdot \left(1-\frac{F}{f_0+F_i}\right)\Bigg].
\end{split}
\label{eqSpecTwo}
\eeq
First, this equation explains the observation that the spectral density of
photocurrents when measuring coherent fields is white (i.e.~frequency
independent). For coherent fields
($t_{01}=t_{10}=0,\,|t_{00}|^2=|t_{11}|^2=1$), the two absolute squares can
be substituted by $|c_\irm|^2$ and the dependence on $F$ cancels. Since in
most practical situations $f_0\gg F$, it is also common to omit the frequency
dependence of the spectrum for squeezed states.

The next step is to include demodulations of the photocurrent. In terms of
the amplitudes defined in Eq.~(\ref{eqCurrAmp}), a singly demodulated
photocurrent, Eq.~(\ref{eqCurrCoh2c1d}), assumes the form
\beq
\hat{I}^{\rm QM}(t)=\frac{1}{2}\sum\limits_{i=0}^{N-1}\int\limits_0^B\drm F\e^{-2\pi\irm F\cdot t}\Big(\hat p(F,c_\irm^*\e^{-\irm\phi},F_\irm-D)+\hat p(F,c_\irm^*\e^{\irm\phi},F_\irm+D)\Big)+\rm h.c.
\eeq
Hereupon, a simple analogy argument leads to the full-fledged multiple
demodulation power spectral density of the photocurrent including squeezed
states of the field. If $C$ frequencies with indices $(n_{\rm j},d_{\rm
j})$, $j\in\{0,..,C-1\}$ of the frequency matrix $\mathcal F$ coincide, then
the respective contribution to the power spectral density is determined by
(compare with Eq.~(\ref{eqSpecNcMd1}))
\begin{widetext}
\beq
\begin{split}
&S^{\rm QM}_{\rm I}(F,\{(n_{\rm j},d_{\rm j})\})=\frac{1}{4^M}\cdot S\left(\sum\limits_{j=0}^{C-1}\hat p(F,c_{n_j}^*\e^{\irm\phi(d_j)},F_{n_0d_0})\right)\\
&\quad = \frac{h\cdot(f_0+F_{n_0d_0})}{2\cdot 4^M}\Bigg[\Bigg|\sum\limits_{j=0}^{C-1}\Big(c_{n_j}^*\e^{\irm\phi(d_j)}t_{00}(f_0+F+F_{n_0d_0})+c_{n_j}\e^{-\irm\phi(d_j)}t_{10}(f_0+F+F_{n_0d_0})\Big)\Bigg|^2\\
&\hspace*{5cm}\cdot\left(1+\frac{F}{f_0+F_{n_0d_0}}\right)\\
&\qquad\qquad\qquad\qquad+\Bigg|\sum\limits_{j=0}^{C-1}\Big(c_{n_j}^*\e^{\irm\phi(d_j)}t_{01}(f_0-F+F_{n_0d_0})+c_{n_j}\e^{-\irm\phi(d_j)}t_{11}(f_0-F+F_{n_0d_0})\Big)\Bigg|^2\\
&\hspace*{5cm}\cdot\left(1-\frac{F}{f_0+F_{n_0d_0}}\right)\Bigg]
\end{split}
\label{eqSpecSQZNM}
\eeq
\end{widetext}
and unique frequencies $F_{\rm nd}$ contribute with
\beq
S^{\rm QM}_{\rm I}(F,(n,d))=\frac{1}{4^M}\cdot S\left(\hat p(F,c_n,F_{nd})\right).
\label{eqSpecGen2}
\eeq
The spectral density on the right-hand side is determined by
Eq.~(\ref{eqSpecTwo}). 

We conclude this section with a brief discussion of
``off-centered'' squeezing, i.e.~some of the squeezing reference frequencies
of the field do not coincide with any of the frequencies $F_{\rm nd}$ or
that squeezing factors $r$ decay over frequency ranges which are comparable
to $F_{\rm nd}\gg B$. We leave a detailed investigation of both problems for the future. One reason is that in each case amplitudes of
the measured field at the photodiode may depend on input amplitudes at more
than two frequencies. In other words, one squeezing process may correlate
amplitudes at say $f_1 = 10^{15}\,{\rm Hz}+50\,{\rm MHz}$ and $f_2 =
10^{15}\,{\rm Hz}+30\,{\rm MHz}$, another squeezing process then correlates
amplitudes at $f_2$ and $f_3 = 10^{15}\,{\rm Hz}-30\,{\rm MHz}$. The output
field at $f_2$ depends on input frequencies $f_1,\,f_2$ and $f_3$. Multiple
correlations are prevented by demanding that the frequency difference of
correlated amplitudes is sufficiently small so that different correlations
occur at well separated parts of the spectrum. If that condition is
fulfilled but ``off-centered'' squeezing is still allowed, then one has to
take greater care when calculating the transfer functions $t_{\rm ij}$. What
if the squeezing reference frequency happens to be at $f_0+F+5\cdot B$ which
is close enough to the measured frequency range $[f_0-B,\,f_0+B]$ to exhibit
some influence on its vacuum noise amplitudes? The answer is derived from
Eq.~(\ref{eqSqueeze}). Let us assume that the squeezing factor and phase are
constant over $[f_0-B,\,f_0+B]$. Then, provided that $B\ll f_0$, the
photocurrent spectral density, Eq.~(\ref{eqSpecSimple}), is modified
according to
\beq
\begin{split}
S^{\rm QM}_{\rm I}(F)&=P_0\cdot hf_0\cdot(\cosh^2(r)+\sinh^2(r))\\
&=P_0\cdot hf_0\cdot\cosh(2r).
\end{split}
\eeq
which means the quantum noise spectral density is necessarily
increased! Therefore, any detector whose sensitivity is limited by optical
quantum noise should avoid off-centered squeezing. 

Again, the previous analysis is not restricted to quantum noise spectra. The
results are valid for any classical, technical noise which is coherent or
exhibits amplitude-phase correlations which can be described by
transformations like Eq.~(\ref{eqTransform}). One may say that these
transformations embody the simplest kind of amplitude-phase correlations. 

\section{Exemplary Calculation}
\label{secExample}

Concluding the paper with an explicit application of our results should be
helpful. We consider the following case: four classical components,
one demodulation and squeezing centered at one of the components $F_{\rm
nd}$ of the frequency matrix $\mathcal F$. The $N=4$ classical components
have frequencies $F_0=0\,\rm Hz$, $F_1=100\,\rm MHz$, $F_2=130\,\rm MHz$ and
$F_3=230\,\rm MHz$. These sideband frequencies are defined with respect to a
large optical frequency $f_0\sim 10^{15}\,\rm Hz$. The photocurrent is
demodulated with $D_0=15\,\rm MHz$. These parameters determine the frequency
matrix
\beq
\mathcal{F}(4,2)=
\begin{pmatrix}
-15\,\rm MHz & 15\,\rm MHz \\
85\,\rm MHz & 115\,\rm MHz \\
115\,\rm MHz & 145\,\rm MHz \\
215\,\rm MHz & 245\,\rm MHz
\end{pmatrix}.
\eeq

We assume squeezing centered around frequency $F_{11}=F_{20}=115\,\rm MHz$
which extends locally over frequencies comparable to the detection bandwidth
$B=1000\,\rm Hz$, all other frequencies contribute coherent vacuum noise. We
already stated that two components of the matrix $\mathcal F$ are equal and
the respective frequency value coincides with the squeezing reference
frequency. Let us first evaluate the contributions from all unique
frequencies. The corresponding vacuum fields are coherent and so the
spectral density Eq.~(\ref{eqSpecTwo}) simplifies to
\beq
\begin{split}
S_{\rm nd}^{\rm coh}&=h(f_0+F_{\rm nd})|c_{\rm n}|^2\\
&=h(f_0+F_{\rm nd})P_{\rm n}.
\end{split}
\eeq
These spectral densities have to be inserted into Eq.~(\ref{eqSpecGen2}) and
summed up for all unique frequencies
\beq
\begin{split}
S_{\rm uni}(F)&=\frac{h}{4}\Big((2f_0+F_{00}+F_{01})P_0+(f_0+F_{10})P_1\\
&\qquad+(f_0+F_{21})P_2+(2f_0+F_{30}+F_{31})P_3\Big)\\
&\approx \frac{hf_0}{4}\Big(2P_0+P_1+P_2+2P_3\Big).
\end{split}
\label{eqSpecEnd1}
\eeq
The remaining problem is to calculate the correlated spectral density at
frequency $F_{11}=F_{20}=115\,\rm MHz$. For simplicity we assume that the
nonlinear transfer is pure squeezing
\beq
\begin{split}
&t_{00}(f_0+F+F_{11})=\cosh(r)\\
&t_{01}(f_0-F+F_{11})=\sinh(r)\e^{\irm 2\phi}\\
&t_{10}(f_0+F+F_{11})=\sinh(r)\e^{-\irm 2\phi}\\
&t_{11}(f_0-F+F_{11})=\cosh(r).
\end{split}
\eeq
Next, we expand the sums in Eq.~(\ref{eqSpecSQZNM}) and substitute all known
parameter values
\beq
\begin{split}
S^{\rm QM}_{\rm I}(F,\{(1,1),(2,0)\})&= \frac{hf_0}{8}\Bigg[\Bigg|(c_1^*\e^{\irm\phi_0}+c_2^*\e^{-\irm\phi_0})\cosh(r)+(c_1\e^{-\irm\phi_0}+c_2\e^{\irm\phi_0})\sinh(r)\e^{-2\irm\phi}\Bigg|^2\\
&\hspace*{1cm}+\Bigg|(c_1^*\e^{\irm\phi_0}+c_2^*\e^{-\irm\phi_0})\sinh(r)\e^{2\irm\phi}+(c_1\e^{-\irm\phi_0}+c_2\e^{\irm\phi_0})\cosh(r)\Bigg|^2\Bigg].
\end{split}
\label{eqSpecTwoEG}
\eeq 
All vacuum energies are approximated by $hf_0$ and $\phi_0$ denotes the
demodulation phase. There are at least two experimentally adjustable phases,
the demodulation phase $\phi_0$ and the squeezing phase $\phi$. What is the
minimum of the spectral density depending on these two phases? Defining
$\alpha\equiv\arg(c_1\e^{-\irm\phi_0}+c_2\e^{\irm\phi_0})$,
Eq.~(\ref{eqSpecTwoEG}) assumes the form
\beq
S^{\rm QM}_{\rm I}(F,\{(1,1),(2,0)\})= \frac{hf_0}{4}|c_1^*\e^{\irm\phi_0}+c_2^*\e^{-\irm\phi_0}|^2\cdot\Big|\cosh(2r)+\sinh(2r)\e^{2\irm(\phi-\alpha)}\Big|^2.
\eeq
Minimization with respect to the squeezing phase is trivial. Setting $\phi^{\rm opt}=\pi/2+\alpha$ and further defining $\Delta\alpha\equiv\arg(c_2)-\arg(c_1)$, the minimized spectral density contribution from the squeezed part of the spectrum simplifies to
\beq
S^{\rm QM}_{\rm I}(F,\{(1,1),(2,0)\})=\frac{hf_0}{4}(P_1+P_2+2\sqrt{P_1P_2}\,\cos(2\phi_0+\Delta\alpha))\cdot\e^{-2r}.
\label{eqSpecExample}
\eeq
Finally, we add this result to Eq.~(\ref{eqSpecEnd1}) which yields
\beq
\begin{split}
S^{\rm QM}_{\rm I}&=\frac{hf_0}{2}\Big(P_0+\frac{1}{2}(1+\e^{-2r})P_1+\frac{1}{2}(1+\e^{-2r})P_2+\sqrt{P_1P_2}\,\cos(2\phi_0+\Delta\alpha)\cdot\e^{-2r}+P_3\Big)\\
&=\frac{hf_0}{2}\Big(P_0+\frac{1}{2}(1+\e^{-2r})P_1+\frac{1}{2}(1+\e^{-2r})P_2-\sqrt{P_1P_2}\cdot\e^{-2r}+P_3\Big).
\end{split}
\eeq
In the last step, we have minimized the noise power by inserting the optimal
demodulation frequency $\phi_0^{\rm opt}=(\pi-\Delta\alpha)/2$. The reader
who is exclusively interested in minimized spectral densities may easily
generalize this final result including squeezing at different frequencies
and an arbitrary number of classical components. The optimization procedure
partly relies on the fact that the squeezing factor and phase are frequency
independent. In general, experimental realization of the smallest possible
noise spectral density requires further degrees of freedom which are
incorporated into the transfer functions $t_{\rm ij}$, i.e.~the light has to
be filtered before the power measurement \cite{KLMTV01,Ha03}. One should
keep in mind that, {\it per se}, a minimal noise spectral density does not
have to be optimal in terms of detector sensitivity. The reason is that
noise minimization simultaneously affects the measured power of the signal.
Sensitivity optimization severely depends on the detector topology
\cite{Ha03}. It was shown in \cite{BCM02} that the sensitivity optimizing
demodulation frequency is $\phi_0^{\rm opt}=-\Delta\alpha/2$ which goes
along with a maximized noise power contribution from
Eq.~(\ref{eqSpecExample}).

\section{Conclusion}

We have presented explicit formulas which govern the power spectral density
of photocurrents generated by power measurements of coherent and squeezed
fields. We have also furnished an appropriate algorithm which can be easily
implemented in quantum-noise simulations. The algorithm is based on a few
limitations concerning the classical spectrum and the squeezed spectrum of
the field. However these limitations are modest and probably insignificant
for most experiments. Our results provide a long-sought-for extension of the Schottky formula
\cite{Sch18,Spe05} to squeezed photon statistics with multiple classical
components. Of special importance is that the shot noise spectrum may be calculated
for any of the currently operating and next generation interferometric 
gravitational wave detectors. Theoretically, the method can be generalized in two ways which we consider to be algorithmically tractable in principle.
First, the transfer functions between different frequencies -- which correspond to classical nonlinearities -- may couple more than three amplitudes at different frequencies. Thereby, multiple squeezing centered around different frequencies with overlapping ranges of non-vanishing squeezing factor could be described. Second, one may want to give up the overlap condition between different ranges of detected field amplitudes. In that case, correlated contributions to the final current spectral density have to be revealed by means of a more elaborate scheme which calculates the intersection boundaries of a partial overlap.\\[0.1cm]

\section{Acknowledgments}

We thank all members of the GEO\,600 Simulation Group \cite{SIM} for helpful
discussions. We also thank the SFB 407 which inspired a great part of this paper. A.~F. would like to thank
PPARC for financial support of this work.

\end{document}